\documentclass[twocolumn,aps,pra,showpacs,amsmath,superscriptaddress]{revtex4}
\usepackage{amssymb}
\usepackage{mathrsfs}
\usepackage{graphicx}
\usepackage{float}

\begin{document}

\title{Wigner crystal induced by dipole-dipole interaction in one-dimensional optical lattices}

\author{Zhihao Xu}
\email{xuzhihao85@gmail.com} \affiliation{Beijing National
Laboratory for Condensed Matter Physics, Institute of Physics,
Chinese Academy of Sciences, Beijing 100190, China}
\author{Shu Chen}
\email{schen@aphy.iphy.ac.cn}\affiliation{Beijing National
Laboratory for Condensed Matter Physics, Institute of Physics,
Chinese Academy of Sciences, Beijing 100190, China}
\date{ \today}

\begin{abstract}
We demonstrate that the static structure factor, momentum
distribution and density distribution provide clear signatures of
the emergence of Wigner crystal for  the fermionic dipolar gas
with strongly repulsive dipole-dipole interactions trapped in
one-dimensional optical lattices. Our numerical evidences are
based on the exact diagonalization of the microscopic effective
lattice Hamiltonian of few particles interacting with long-range
interactions. As a comparison, we also study the system with only
nearest-neighbor interactions, which displays quite different
behaviors from the dipolar system in the regime of strong
repulsion.
\end{abstract}


\maketitle

\section{Introduction}
Ultracold dipolar atomic and molecular gases with long-range
interactions have become a very active research field of cold atom
physics in past years. Dipolar interactions with long-range
anisotropic character have been observed in Chromium atoms
\cite{Thierry Lahaye,Stuhler}. More recently, the development of
stimulated Raman adiabatic passage technique has succeeded in
creating a nearly degenerate gas of heteronuclear polar molecule
gases which have much greater dipole momentum \cite{KK
Ni,Ospelkaus,D Wang} and thus are promising candidates with very
strong dipolar interactions. The experimental progress has
stimulated theoretical studies of dipolar gases with long-range
anisotropic interactions. Loading ultracold polar molecules into
optical lattices also provides a fascinating platform for studying
quantum many-body systems with long range interactions.

In comparison with the short-range and isotropic interactions,
dipole-dipole interactions (DDIs) may induce many new effects and
phases in quantum gases due to their long-range and anisotropic
character. There have been many theoretical works on, for example,
the polarized dipolar Bose-Einstein condensations (BECs) \cite{S
Yi1,Santos1,Muller}, spinor-dipolar BECs
\cite{Kawahuchi,Santos2,J. -N. Zhang}, supersolid
\cite{Goral,Trefzger,Sansone,Golomedov,S Yi2}, s-wave scattering
resonances\cite{Zhe-Yu Shi}, quantum Hall effects \cite{R.-Z. Qiu}
and one-dimensional (1D) dipolar systems
\cite{Sinha,Citro,Zollner,Deurezbacher,Arkhipov}. In the presence
of long range interaction, an interesting issue is the emergence
of the Wigner crystal phase. A Wigner crystal is a crystalline
phase of electrons first predicted by Wigner \cite{Wigner}. As the
interacting potential energy dominates the kinetic energy at low
densities, the particles tend to form a regular crystal due to the
strong long-range interaction. Wigner crystal in one dimension
with long range interactions has been theoretically studied by
using different methods, including bosonization
\cite{Schulz,Inoue,Tsukamoto,Dalmonte}, quantum Monte Carlo (QMC)
\cite{Lee,Astrakharchik,Citro,Arkhipov} and exact diagonaliztion
\cite{Deurezbacher}. In the seminal work of Schulz \cite{Schulz},
it has been clarified that the Coulomb long-range repulsion
enhances the 4$k_F$ charge density correlations and drives the
system to the Wigner crystal phase.

In this paper, we study the fermionic dipolar gas with strong
dipole-dipole interactions trapped in 1D deep optical lattices,
which can be effectively described by a Fermi lattice model with
long-range interactions falling off as $1/x^3$. For 1D fermionic
models, the case of interactions falling off as $1/x^{\alpha}$ has
been studied in the scheme of bosonization \cite{Inoue,Tsukamoto}.
Treating long-range forward scattering as a perturbation, they
find that for $\alpha>1$ the long-range forward scattering is an
irrelevant perturbation.  As the above conclusion is obtained
based on the perturbation analysis of the low-energy effective
Luttinger liquid theory with linear dispersion, it does not
exclude the existence of Wigner crystal phase in the presence of
strong dipole-dipole interactions. In this work, we shall study
the ground state properties of the Fermi lattice model with
long-range interactions by means of the exact diagonalization
method. To see clearly the effect of the long-range interaction,
we carefully compare the long-range interacting dipolar systems
against the Fermi systems with short-range interactions.  Our
results display significantly different behaviors for the Fermi
systems with long-range and short-range interactions in the
strongly interacting regime. By observing the differences of
static structure factor, momentum distribution and density
distribution, we give clear evidences for the emergence of Wigner
crystal for dipolar fermions with strong dipole-dipole interactions.

The paper is organized as follows. In section II, we derive the
effective lattice model for the fermionic dipolar gas in a 1D
optical lattice. In section III, we present our results for both
the dipolar Fermi system with long range interaction and system
with only nearest-neighbor (NN) interaction  by using the exact
diagonalization method and compare their different behaviors with
increasing the interaction strength. A summary is given in the
last section.

\section{Model Hamiltonian}
Let us consider a system of electric or magnetic dipoles confined
in a quasi-1D optical lattice along the $x$ direction. The
interaction between dipoles aligned in the $x-z$ plane by a
homogeneous filed can be simplified to
\begin{equation}
\label{eqn1}
V_{dd}(\vec{r})=C_{dd}\frac{1-3\cos^2{\theta_{rd}}}{r^3},
\end{equation}
where $\cos{\theta_{rd}}=\vec{r}\cdot\vec{d}/(rd)$ and $C_{dd}$
measures the strength of the DDI. The strength of the DDI is given
by $C_{dd}=d^2/(4\pi\varepsilon_0)$ for two electric dipoles, and
by $C_{dd}=\mu_0d^2/(4\pi)$ for the magnetic ones, with
$\varepsilon_0$ and $\mu_0$ being the electric constant and
magnetic constant, respectively. In the single-mode approximation
with only the transversal ground state considered, we can
integrate over transversal directions and get the effective 1D DDI
given by $V_{dd}=U_{dd}\widetilde{V}_{dd}(|x|/l_{\perp})$, with
$U_{dd}=-C_{dd}[1+3\cos(2\theta)]/(8l_{\perp}^3)$,
$\widetilde{V}_{dd}(u)=-2u+\sqrt{2\pi}(1+u^2)\exp(u^2/2)\mathrm{erfc}(u/\sqrt{2})$
and $l_\perp=\sqrt{\hbar/(m\omega_{\perp})}$
\cite{Deurezbacher,Sinha}.

The Hamiltonian for dipolar fermions in optical lattices is given by
\begin{eqnarray}
\label{eqn2}
H&=&\int \mathrm{d}x \psi^{\dagger}(x)\left[-\frac{\hbar^2}{2m}\frac{\partial^2}{\partial x^2}+V_0(x)\right]\psi(x)\\
&&+\frac{1}{2}\int \int \mathrm{d}x \mathrm{d}x^{\prime} \psi^{\dagger}(x)\psi^{\dagger}(x^{\prime})V_{dd}(x-x^{\prime})\psi(x^{\prime})\psi(x),\notag
\end{eqnarray}
where $m$ is the mass of the dipolar fermions, $\psi(x)$ is a
fermionic field operator for the dipolar fermion, and $V_0(x)$ is
the optical lattice field given by $V_0(x)=V_0\sin^2(kx)$ with the
wavevectors $k=2\pi/\lambda$ and the wavelength of the laser light
$\lambda$. Considering the deep optical lattice with particles
trapped in the lowest vibrational state
$\omega(x)=\exp(-x^2/(2l_x^2))/(\pi^{1/4}\sqrt{l_x})$ with
$l_x=\sqrt{\hbar/(m\omega_x)}$ and $\hbar \omega_x = 2 \sqrt{E_R
V_0}$,
we use the Wannier basis to expand the field operator
$\psi(x)=\sum_ic_i\omega(x-x_i)$. Similar to the case of Hubbard
model \cite{Jaksch}, we can get the effective lattice Hamiltonian
for the polar fermionic system
\begin{equation}
\label{eqn3}
H=-J\sum_i(c^\dagger_ic_{i+1}+\mathrm{H.c.})+\frac{1}{2}\sum_{i\neq j}V_{dd}(|i-j|)n_in_j,
\end{equation}
where $c^\dagger_i(c_i)$ is the creation (annihilation) operator
of the fermion, $J=\sqrt{E_RV_0}\exp(-\pi^2\sqrt{V_0/E_R}/4)$ with
$E_R=\hbar^2 k^2/2m$ the recoil energy of the system,
$V_{dd}(|i-j|)=U_{dd} \int \int\mathrm{d}x \mathrm{d}x^{\prime}
\widetilde{V}_{dd}(|x-x^{\prime}|/l_{\perp})
 \omega^*(x-x_i)\omega^*(x^{\prime}-x_j)\omega(x-x_i)\omega(x^{\prime}-x_j)$
is the DDI between the dipoles at the positions of $x_i$ and
$x_j$.

\begin{figure}[tbp]
\includegraphics[height=9cm,width=9cm] {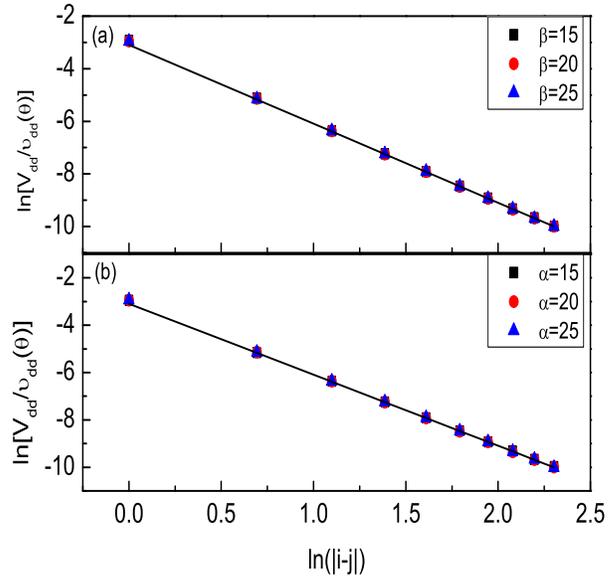}
\caption{(Color online) Log-Log plot of
$V_{dd}/\upsilon_{dd}(\theta)$ versus distances between two
dipoles in optical lattices for different $\alpha$ and $\beta$
with $\upsilon_{dd}(\theta)=-C_{dd}[1+3\cos(2\theta)]$. (a) We set
$\alpha=20$ for different $\beta$, and the slop of the solid line is $-3$;
(b) $\beta=20$ is fixed for $\alpha=15$, $20$, $25$, also the gradient of the fitting line
is $-3$.
}\label{Fig1}
\end{figure}

After the numerical calculation by setting $E_R$, $\hbar$, and $m$
as units, $\omega_{\perp}/\omega_x=\alpha$ and $V_0/E_R=\beta$,
the $V_{dd}(|i-j|)$ is found to decay as $1/|i-j|^3$ according to
Fig.\ref{Fig1}. In Fig. \ref{Fig1}, we show the Log-Log plot of
$V_{dd}/\upsilon_{dd}(\theta)$ versus distances between dipoles in
1D optical lattices for different $\alpha$ and $\beta$ with
$\upsilon_{dd}(\theta)= - C_{dd}[1+3\cos(2\theta)]$, where
$\theta$ is the angle between the dipole direction and $x$ axis.
When $\alpha$ is fixed for different $\beta$ (see
Fig.\ref{Fig1}(a)), the gradient of the fitting line is $-3$. In
the Fig.\ref{Fig1}(b), the slops of the fitting line is the same
as the one in the Fig.\ref{Fig1}(a) where $\beta$ is fixed and
$\alpha$ is changing. Also we can see when $\alpha$ is large
enough, $V_{dd}$ changes little as $\alpha$ increases, because all
the dipoles have been confined along the axial direction and can
be seen as a one dimensional system. So we can simplify the
Hamiltonian (\ref{eqn3}) as
\begin{equation}
\label{eqn4}
H=-J\sum_i(c^\dagger_ic_{i+1}+\mathrm{H.c.})+\frac{1}{2}V\sum_{i\neq j}\frac {n_in_j}{|i-j|^3},
\end{equation}
with $V$ the strength of the DDI. For comparison, we also consider
the system with only short-range interactions described by the
following hamiltonian
\begin{equation}
\label{eqn5}
H=-J\sum_i(c^\dagger_ic_{i+1}+\mathrm{H.c.})+\frac{1}{2}V\sum_{
\langle i,j \rangle} n_in_{j},
\end{equation}
where only the nearest-neighbor interaction is considered and
$\langle i,j \rangle$ means summation over nearest neighbors. In
the present work, we only consider the case with repulsive
interaction $V>0$. For simplicity, we take $\theta = \pi/2$.

\section{Results}

We shall study the ground state properties for both systems
described by Hamiltonians (\ref{eqn4}) and (\ref{eqn5}) by the
exact diagonalization method. To give a concrete example, we focus
on the system with five particles ($N=5$) in a lattice with size
$L=30$ under the periodic boundary condition. In order to gain
some intuitive understanding of effect of long-range interactions,
we first consider the limiting cases before presenting our
calculated results. In the limit of $V/J \to 0$, the hopping term
dominates and prevents the formation of crystal phase, and thus
the difference of the long-range interaction and nearest-neighbor
interaction is not obvious. However, in the strongly interacting
limit of $V/J \to \infty$, the hopping term can be ignored, and
the effect of long-range interaction becomes significant. The
long-range repulsive interaction tends to repel particles to be
equally spaced and form a solid like state. When apart from the
strongly interacting limit, the hopping processes and quantum
fluctuations prevent the formation of a perfect Wigner crystal.

\begin{figure}[tbp]
\includegraphics[height=9cm,width=9cm] {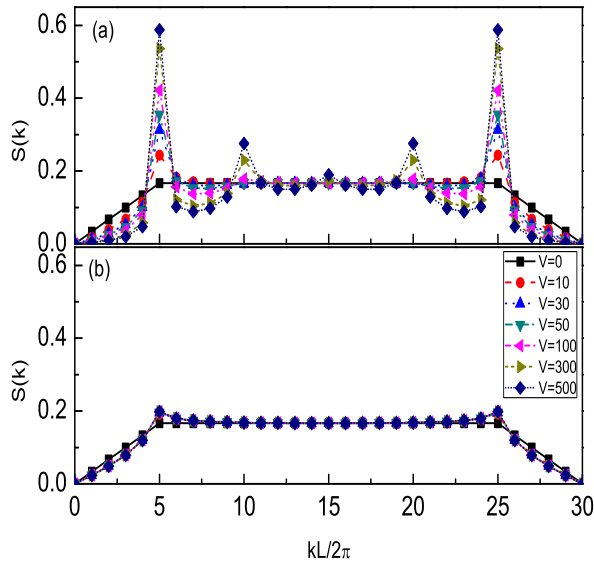}
\caption{(Color online) Static structure factors $S(k)$ for
systems with $L=30$, $N=5$ and different $V$. (a) The dipolar
Fermi system, and (b) the Fermi system with only the
nearest-neighbor interaction. }\label{Fig2}
\end{figure}
\begin{figure}[tbp]
\includegraphics[height=9cm,width=9cm] {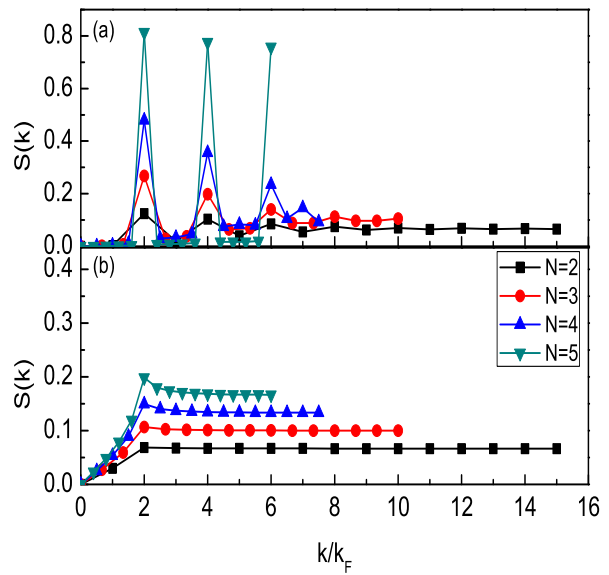}
\caption{(Color online) Static structure factors $S(k)$ vs
$k/k_F$ for systems with $L=30$, $V=5000$ and different $N$.
(a) is for the dipolar system, and (b) is for system with short
range interaction. }\label{Fig3}
\end{figure}

To characterize the phase of Wigner crystal, we calculate the
static structure factor, defined as
\begin{equation}
\label{eqn6} S(k)=\frac{1}{L}\sum_{i,j}e^{ik(i-j)}[\langle n_i n_j
\rangle - \langle n_i \rangle \langle n_j \rangle],
\end{equation}
where $k=2m\pi/L$ with $m=0,1,...,L$. The phase of Wigner crystal
can be characterized by the onset of the $4k_F$ peak in the static
structure factor with $k_F=n\pi$ and $n=N/L$. When $V$ is small, we do not find obvious
differences of $S(k)$ for systems with either long-range or
short-range interaction. However, significant differences are
emergent as the interaction strength increases to the strongly
interacting limit. In Figure \ref{Fig2}, we display the static
structure factor $S(k)$ for different values of $V=0,10,50,100$
and $500$. For both systems, there is a peak at $k=2k_F$ for
$V=10$ despite that the peak for the system with long-range
interaction is more obvious. As the interaction increases further,
more peaks emerge at reciprocal lattice vectors $kL/2\pi N =
integer$ for the system with the long-range interaction
(Fig.{\ref{Fig2}}a). The
height of peak increases with increasing $V$, evidencing the
occurrence of the Wigner crystal state.
While for the system with short
range interaction described by Hamiltonian (\ref{eqn5}), from
Fig.\ref{Fig2}(b) we can see that only peaks at $\pm 2k_F$ occur
and there are no other peaks appearing.  Also, the height of peaks
at $\pm 2 k_F$ increases more slowly with the increase of $V$ in
contrast to the system with dipole-dipole interaction.

We also demonstrate data of static structure factor $S(k)$ versus
$k/k_F$ in Fig.\ref{Fig3} for systems with $L=30$, $V=5000$
and $N=2,3,4,5$.
Systems with different $N$ display similar behaviors in the large
$V$ limit. As shown in Fig.\ref{Fig3}(a), for systems described by
Eq.(\ref{eqn4}) with different filling factors, peaks of $S(k)$
emerge at reciprocal lattice vectors $k/k_F=2m$ with $m$ the
integer. The appearance of peaks for various filling cases
indicates that the Wigner crystal emerges when the filling is
either commensurate or incommensurate with the optical lattice.
For the systems described by Eq.(\ref{eqn5}), Fig.\ref{Fig3}(b)
shows that there are no other peaks except at $k/k_F=2$.
\begin{figure}[tbp]
\includegraphics[height=8cm,width=8cm] {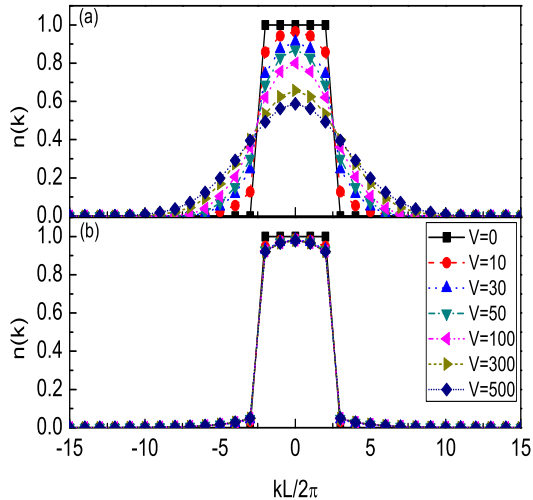}
\caption{(Color online) Momentum distributions for systems with
$L=30$, $N=5$ and different $V$. (a) The dipolar Fermi system, (b)
the Fermi system with nearest-neighbor interaction.}\label{Fig4}
\end{figure}
\begin{figure}[tbp]
\includegraphics[height=8cm,width=8cm] {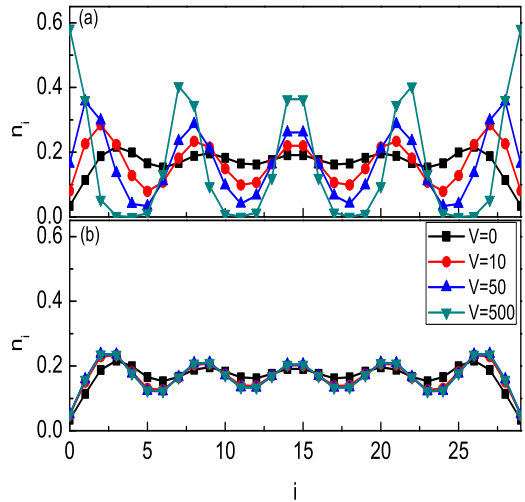}
\caption{(Color online) Density distributions for systems with
$L=30$, $N=5$ and different $V$ under the open boundary condition.
(a) The dipolar Fermi system, and (b) the Fermi system with
nearest-neighbor interaction. }\label{Fig5}
\end{figure}

Next we calculate the momentum distribution, which is defined by
the Fourier transform with respect to $i-j$ of the one-particle
density matrix with the form
\begin{equation}
\label{nk} n(k)=\frac{1}{L}\sum_{i,j}e^{ik(i-j)} \langle
c^{\dagger}_i c_j \rangle .
\end{equation}
In Fig.\ref{Fig4}, we show momentum distributions for both systems
with different $V$. Fig.\ref{Fig4}(a) is corresponding to the
dipolar model, and Fig.\ref{Fig4}(b) is for the model with NN
interactions. We also find the similarity of the two different
models in the regime of small $V$. They show typical momentum step
distribution of fermions. But with increasing $V$, the momentum
densities of the dipolar model become fatter and lower
(Fig.\ref{Fig4}(a)). At last for very strong $V$, the momentum
distribution displays the Gaussian distribution which mainly comes
from the Fourier transformation of the diagonal part of the
density matrix, revealing that there is no overlap between the
localized wave packets with the non-diagonal density matrix
tending to vanish. In contrast to the case with DDI, the momentum
distributions in Fig.\ref{Fig4}(b) change not obviously even for
very strong $V$.

To gain an intuitive insight on the crystalline phase, we analyze
the density distribution $n_i = \langle c^{\dagger}_i c_i \rangle
$ for systems described by Hamiltonian (\ref{eqn4}) and
(\ref{eqn5}) with open boundary conditions. In Fig. \ref{Fig5} we
show density distributions of five particles for different $V$.
As shown in Fig.\ref{Fig5}(a) for the DDI systems, with increasing
the long-range interaction, particles move apart each other with
more sharp peaks emerging in the density profile. When $V$ is
large enough, the density resembles five well-separated localized
wave packs, which are totally apart from each other with the
equilibrium positions of wave packets minimizing the interaction
energy, characterizing the system evolved into the Wigner crystal
regime \cite{Deurezbacher}. As a comparison, the density
distributions shown in Fig.\ref{Fig5}(b) only change marginally
with increasing the nearest-neighbor repulsion. No crystalline
signal is detected even in the limit of strong repulsion.

Before ending this paper, we would like to discuss the validity of
the approximation of the lowest transverse state in deriving the
effective 1D Hamiltonian (\ref{eqn2}). Except of the requirement
of $\omega_{\perp}/\omega_{x} \gg 1$, one also needs that the
interacting energy $E_{DDI}$ induced by the dipolar interaction is
much smaller than the energy of transverse confinement $\hbar
\omega_{\perp}$. Different from the case with contact interaction,
the interacting energy of the dipolar system is proportional to
the interaction strength $V$, and thus the approximation is
expected to break down if $E_{DDI} > \hbar \omega_{\perp}$. For
the present case, the longitudinal kinetic energy is greatly
suppressed due to the existence of the longitudinal optical
lattice, i.e., $J \ll \hbar \omega_x$. Consequently, even when
$V/J \gg 1$, the interacting energy is still much smaller than
$\hbar \omega_{\perp}$ and thus the effective 1D description still
holds true even in the regime of formation of Wigner crystal.
Taking the case corresponding to Fig.2 as an example, if we take
$\alpha=20$ and $\beta=20$, $J/\hbar \omega_x \sim 10^{-6}$, and
therefore even for $V/J=500$ the $E_{DDI}$ is still much smaller
than $\hbar \omega_{\perp}$. For systems with even stronger
interaction or with large atom numbers, one can also tune $\alpha
\gg 1$ to fulfill the requirement of $E_{DDI} \ll \hbar
\omega_{\perp}$. In the future work, it would be also interesting
to study the crossover from 1D to high-dimensional system when the
transverse confinement decreases, for which the approximation of
the lowest transverse state does not work well and new phenomena
may appear \cite{Cooper}.

\section{Summary}
In summary, we have studied the dipolar fermionic system with
dipole-dipole interactions trapped in 1D optical lattices by means
of the exact diagonalization method. We have shown that the static
structure factor, momentum distribution and density distribution
provide clear evidences for the existence of Wigner crystal, as
the interaction energy overcomes the energy scale of hopping
energy. We also compare our results of the dipolar system to the
model with only nearest neighbor interactions, which exhibits no
signatures of Wigner crystal even in the strongly interacting
limit. Our study unveils the important role of the long range
interaction in the formation of the Wigner crystal.

\begin{acknowledgments}
We thank X. L. Gao and S. Yi for helpful discussions. This work
was supported by the 973 projects of the Ministry of Science and
Technology of China (2011CB921700), NSF of China under Grants No.
10821403, No. 11174360 and No. 10974234.
\end{acknowledgments}


\begin{references}

\bibitem{Thierry Lahaye} T. Lahaye, T. Koch,
B. Fr\"{o}lich, M. Fattori, J. Metz, A. Griesmaier, S. Giovanazzi,
and T. Pfau, Nature \textbf{448}, 672 (2007).

\bibitem{Stuhler} J. Stuhler, A. Griesmaier, T. Koch,
 M. Fattori, T. Pfau, S. Giovanazzi, P. Pedri, and
 L. Santos, Phys. Rev. Lett \textbf{95},
150406 (2005).

\bibitem{KK Ni} K.-K. Ni, S. Ospelkaus, M. H. G. de Miranda,
A. Pe'er, B. Neyenhuis, J. J. Zirbel, S. Kotochigova, P. S. Julienne,
D. S. Jin, and J. Ye, Science \textbf{322},
231 (2008).

\bibitem{Ospelkaus}
S. Ospelkaus, A. Pe'er, K.-K. Ni, J. J. Zirbel, B. Neyenhuis, S.
Kotochigova, P. S. Julienne, J. Ye, and D. S. Jin, Nat. Phys.
\textbf{4}, 622 (2008).

\bibitem{D Wang} D. Wang, B. Neyenhuis, M. H. G. de Miranda, K.-K. Ni, S. Ospelkaus,
D. S. Jin, and J. Ye, Phys. Rev A \textbf{81}, 061404 (2010).

\bibitem{S Yi1} S. Yi and L. You, Phys. Rev. A \textbf{61},
041604(R) (2000); Phys. Rev. A \textbf{63}, 053607 (2001).

\bibitem{Santos1} L. Santos, G.V. Shlyapnikov, P. Zoller,
and M. Lewenstein, Phys. Rev. Lett. \textbf{85}, 1791 (2000).

\bibitem{Muller} S. M\"{u}ller, J. Billy, E. A. L. Henn,
H. Kadau, A. Griesmaier, M. Jona-Lasinio, L. Santos, and
T. Pfau, Phys. Rev. A \textbf{84},
053601 (2011).

\bibitem{Kawahuchi} Y. Kawaguchi, H. Saito, and M. Ueda,
Phys. Rev. Lett. \textbf{96}, 080405 (2006).

\bibitem{Santos2} L. Santos and T. Pfau, Phys. Rev. Lett. \textbf{96},
190494 (2006).

\bibitem{J. -N. Zhang} J.-N. Zhang, L. He, H. Pu, C.-P. Sun, and S. Yi,
Phys. Rev. A \textbf{79},
033615 (2009).

\bibitem{Goral} K. G\'{o}ral, L. Santos, and M. Lewenstein, Phys. Rev. Lett. \textbf{88},
170406 (2002).



\bibitem{Trefzger} C. Trefzger, C. Menotti, and M. Lewenstein, Phys. Rev. Lett. \textbf{103},
035304 (2009).

\bibitem{Sansone} B. Capogrosso-Sansone, C. Trefzger, M. Lewenstein, P. Zoller,
and G. Pupillo, Phys. Rev. Lett. \textbf{104}, 125301 (2010).

\bibitem{Golomedov} A. E. Golomedov, G. E. Astrakharchik, and Yu. E. Lozovik,
Phys. Rev. A \textbf{84},
033615 (2011).

\bibitem{S Yi2} S. Yi, T. Li, and C. P. Sun, Phys. Rev. Lett. \textbf{98},
260405 (2007).

\bibitem{Zhe-Yu Shi} Z.-Y. Shi, R. Qi and H. Zhai, arXiv:1108.3510.

\bibitem{R.-Z. Qiu} R. -Z. Qiu, S.-P. Kou, Z.-X. Hu, X. Wan, and S. Yi,
Phys. Rev. A \textbf{83},
063633 (2011).

\bibitem{Sinha} S. Sinha and L. Santos, Phys. Rev. Lett. \textbf{99},
140406 (2007).


\bibitem{Arkhipov} A. S. Arkhipov, G. E. Astrakharchik, A. V. Belikov, and Y. E.
Lozovik, JETP Lett. 82, 39 (2005).


\bibitem{Citro} R. Citro E. Orignac, S. De Palo, and M. L. Chiofalo,
Phys. Rev. A \textbf{75}, 051602(R) (2007).

\bibitem{Deurezbacher} F. Deurezbacher, J. C. Cremon, and S. M. Reimann, Phys. Rev. A \textbf{81},
063616 (2010).

\bibitem{Zollner} S. Z\"{o}llner, G. M. Bruun, C. J. Pethick, and S. M. Reimann, Phys. Rev. Lett. \textbf{107},
035301 (2011).

\bibitem{Wigner} E. Wigner, Phys. Rev. \textbf{46}, 1002 (1934).

\bibitem{Schulz} H. J. Schulz Phys. Rev. Lett. \textbf{71},
1864 (1993).

\bibitem{Inoue} H. Inoue and K. Nomura, J. Phys. A: Math. Gen. \textbf{39}, 2161
(2006).

\bibitem{Tsukamoto} Y. Tsukamoto and N. Kawakami, J. Phys. Soc. Jpn. 69, 149
(2000).

\bibitem{Dalmonte} M. Dalmonte, G. Pupillo, and P. Zoller, Phys. Rev. Lett. \textbf{105}, 140401
(2010).

\bibitem{Astrakharchik} G. E. Astrakharchik and M.D. Girardeau, Phys. Rev. B \textbf{83},
153303 (2011).

\bibitem{Lee} R. M. Lee and N. D. Drummond, Phys. Rev. B \textbf{83},
245114 (2011).





\bibitem{Jaksch} D. Jaksch, C. Bruder, J. I. Cirac, C. W. Gardiner, and P. Zoller,
Phys. Rev. Lett. \textbf{81}, 3108 (1998).

\bibitem{Cooper} See, for example, S. Komineas and N. R. Cooper, Phys. Rev. A \textbf{75}, 023623 (2007).



\end{references}
\end{document}